\documentclass[a4paper,11pt]{article}
\usepackage{jcappub} 
\usepackage{aas_macros}
\usepackage{graphicx,color}
\graphicspath{{Figures/}}
\usepackage{subcaption}
\usepackage[dvipsnames]{xcolor}
\usepackage{amsmath}
\usepackage{booktabs}
\usepackage{hyperref}
\usepackage[normalem]{ulem}
\usepackage[T1]{fontenc} 
\usepackage{capt-of}
\usepackage{multirow}
\usepackage{siunitx}
\ifdefined\qty\else
  \ifdefined\NewCommandCopy
    \NewCommandCopy\qty\SI
  \else
    \NewDocumentCommand\qty{O{}mm}{\SI[#1]{#2}{#3}}
  \fi
\fi
\ifdefined\unit\else
  \ifdefined\NewCommandCopy
    \NewCommandCopy\unit\si
  \else
    \NewDocumentCommand\unit{O{}m}{\si[#1]{#2}}
  \fi
\fi

\title{SECRET: Stochasticity Emulator for Cosmic Ray Electrons}

\author[a,b]{Nikolas Frediani,}
\author[a]{Michael Kr\"amer,}
\author[a]{Philipp Mertsch}
\author[a]{and Kathrin Nippel}

\affiliation[a]{Institute for Theoretical Particle Physics and Cosmology, RWTH Aachen University, D-52056 Aachen, Germany}
\affiliation[b]{University Observatory, Ludwig-Maximilians University Munich, D-81679 Munich, Germany}

\emailAdd{Nikolas.Frediani@physik.lmu.de}
\arxivnumber{TTK-25-02}

\abstract{ 
The spectrum of cosmic-ray electrons depends sensitively on the history and spatial distribution of nearby sources. 
Given our limited observational handle on cosmic-ray sources, any model remains necessarily probabilistic. 
Previously, predictions were performed in a Monte Carlo fashion, summing the contributions from individual, simulated sources to generate samples from the statistical ensemble of possible electron spectra. 
Such simulations need to be re-run if the cosmic-ray transport parameters (e.g. diffusion coefficient, maximum energy) are changed, rendering any parameter study computationally expensive. 
In addition, a proper statistical analysis of observations and comparison with such probabilistic models requires the joint probability distribution of the full spectrum instead of only samples. 
Note that parametrising this joint distribution is rendered difficult by the non-Gaussian statistics of the cosmic-ray fluxes. 
Here, we employ machine learning to compute the joint probability distribution of cosmic-ray electron fluxes. 
Specifically, we employ a Masked Autoencoder for Distribution Estimation (MADE) for a representation of the high-dimensional joint probability distribution. 
In a first step, we train the network on a Monte Carlo simulation for a fixed set of transport parameters, thus significantly accelerating the generation of samples. 
In a second step, we extend this setup to SECRET (Stochasticity Emulator for Cosmic Ray Electrons), allowing to reliably interpolate over the space of transport parameters. 
We make the MADE and SECRET codes available at \href{https://git.rwth-aachen.de/pmertsch/secret}{this URL}. 
}

\begin{document}

\maketitle


\section{Introduction}


Despite intense theoretical and observational progress over past decades, the origin of cosmic rays (CRs) is still an open and pressing problem~\cite{Gabici:2019jvz}. 
At energies below $1 \, \text{PeV}$, there is agreement that sources must galactic~\cite{ParticleDataGroup:2024cfk}. 
Identifying individual sources is hampered by the fact that as charged particles, galactic CRs travel diffusively through the turbulent galactic magnetic fields, such that the observed arrival directions do not point back to the sources.
At the same time, if the sources of galactic CRs are supernova remnants (SNRs) as commonly assumed~\cite{1964ocr..book.....G}, the number of them that significantly contribute to the intensity of CRs on Earth is very large. 
Over the time-scales of tens of mega-years during which GeV CRs are confined to the CR halo, tens of thousands of supernova explosions occur within a distance of a few kiloparsec. 
The contribution of any such source to the total intensity is therefore very small and so are spectral features from individual sources. 

The situation is however different for CR electrons (CREs)\footnote{Here and in the following we take CR electrons to mean the sum of electrons and positrons.}. 
While they also travel diffusively, radiative losses severely limit their range and hence the number of sources that contribute at any one position. 
The relevant time scale for CREs above a few GeV is not the diffusive escape time but rather the energy loss time which can be approximated as $2 \times 10^2 \, \text{Myr} \, (E/\text{GeV})^{-1}$. 
For typical values of the diffusion coefficient, $\kappa \simeq 3 \times 10^{28} \text{cm}^2/\text{s} \simeq 0.1 \, \text{kpc}^2/\text{Myr}$ and $1 \, \text{kpc}^2/\text{Myr}$ at $1 \, \text{GeV}$ and $1 \, \text{TeV}$, respectively, this results in spatial ranges of $\sim 4 \, \text{kpc}$ and $\sim 0.1 \, \text{kpc}$. 
For the same source rate, the number of sources contributing significantly is therefore $\mathcal{O}(10^4)$ and $\mathcal{O}(1)$, again at $1 \, \text{GeV}$ and $1 \, \text{TeV}$, respectively. 
Beyond $\sim 1 \, \text{TeV}$, we therefore expect spectral features of individual sources to occur in the spectra of CREs. 
Observations of CREs at TeV energies therefore offer a great opportunity for identifying the sources of CRs. 

The intensity $\phi$ of CREs, defined as the number of particles per unit area, time, solid angle and energy has been measured with great precision by a number of experiments\footnote{For a recent compilation of data, see e.g. Fig.~30.4 of Ref.~\cite{ParticleDataGroup:2024cfk}.}. 
At energies below $\sim 10 \, \text{GeV}$ the intensity of CREs is markedly affected by solar modulation~\cite{Potgieter:2013pdj,Vittino:2019yme,Kuhlen:2019hqb}. 
At higher energies, the spectrum roughly follows a power law $J(E) \propto E^{-3.1}$, although with a number of noteworthy features: 
At $\sim 30 \, \text{GeV}$, the spectrum hardens, that is the the spectrum starts to decrease more slowly with energy. 
This is also the energy where discrepancies between different experiments become significant. 
While below $\sim 30 \, \text{GeV}$, measurements by the four most recent space experiments agree, above $\sim 30 \, \text{GeV}$ the intensities measured by DAMPE~\cite{DAMPE:2017fbg} and \emph{Fermi}-LAT~\cite{Fermi-LAT:2017bpc} are higher than those measured by AMS-02~\cite{AMS:2021nhj} and CALET~\cite{Adriani:2018ktz}. 
As for the interpretation of the $\sim 30 \, \text{GeV}$ hardening, it has been claimed~\cite{Evoli:2020ash,Fang:2020dmi} that this is the result of a change in the cooling rate, due to the Klein-Nishina suppression of inverse-Compton scattering on photons of optical frequencies. 
Others have argued that this effect is not very strong and instead the break could be explained by a new population of sources with a harder spectrum starting to contribute~\cite{DiMauro:2020cbn}. 
While the $30 \, \text{GeV}$ break is rather subtle, a very prominent feature is a break at around $1 \, \text{TeV}$ where the spectrum softens by about one power in energy. 
This was first observed by H.E.S.S.~\cite{HESS:2008ibn} and later confirmed by a number of other experiments~\cite{BorlaTridon:2011dk,Staszak:2015kza,DAMPE:2017fbg}. 
The latest analysis by H.E.S.S.~\cite{2024PhRvL.133v1001A} that extends to energies as high as $15 \, \text{TeV}$, finds that the spectral index changes from $3.25$ to $4.49$ with a break energy of $1.17 \, \text{TeV}$. 
This break has been interpreted as a feature from an individual source which would need to be dominating at energies around the break~\cite{Recchia:2018jun,Fornieri:2020kch} or as the feature from a statistical ensemble of sources~\cite{Mertsch:2018bqd}. 

In many phenomenological models of the CRE spectrum~\cite{Gaggero:2013rya,Boudaud:2016jvj,Boschini:2018zdv,Vittino:2019yme,Tian:2019bfy,Diesing:2019lwm,Evoli:2020ash}, however, effects due to the stochasticity of sources are neglected. 
Instead the distribution of sources are approximated with a smooth function of position. 
This leads to a prediction for the CRE intensity that is deterministic. 
However, at energies where stochasticity effects are important, that is at hundreds of GeV and above, neglecting the stochasticity can lead to faulty conclusion, for instance when parameters are inferred by fitting to data. 

As for models that do consider stochasticity effects, three approaches can be distinguished: 
Some models~\cite{Kobayashi:2003kp,Recchia:2018jun,Fornieri:2020kch} focus on particular spectral features, which they try to ascribe to individual sources, on top of a background of a smooth distribution of sources. 
This can run into the danger of over- or undercounting of sources~\cite{Mertsch:2018bqd}. 
Other models~\cite{Malyshev:2009tw,DiMauro:2017jpu,DiMauro:2019yvh,DiMauro:2020cbn} employ source positions and ages from catalogues of, e.g. SNRs or pulsars as proxies for the likely position of CRE sources. 
The catalogues employed are however necessarily incomplete such they are oftentimes complemented by a smooth distribution of far away sources. 
This can lead to an underestimate in nearby, but old sources which are typically not present in catalogues~\cite{Mertsch:2018bqd}. 
Ultimately, only purely stochastic models~\cite{Pohl:1998ug,Blasi:2010de,Mertsch:2010fn,Cholis:2017ccs,Mertsch:2018bqd,Porter:2019wih,Evoli:2020szd,Evoli:2021ugn,Cholis:2021kqk,Orusa:2021tts,Asano:2021wze,Marinos:2024rcg} can provide a consistent description of CRE intensities. 
In these models, a distribution of sources is drawn from a probability density and the intensities for the simulated sources are summed up. 
Repeating this Monte Carlo (MC) procedure for different source configurations, allows building a statistical ensemble which can be analysed statistically. 

As the latter approach requires simulating a large number of ensemble members to bring down sample variance in the analysed quantities, it is computationally rather expensive. 
However, analytical solutions for even only the marginal distribution of the CRE intensity at individual energies are only available under simplifying assumptions and in certain limits~\cite{Mertsch:2010fn}. 
In addition, only considering marginal distributions neglects the correlation of the CRE intensities between different energies which is the key to identifying individual sources. 
Instead, what is needed is the joint distribution of the spectrum at different energies. 
Previously, this has been addressed by approximating the joint distribution with a copula construction, with the copula parameters determined by fitting to the results of MC simulations~\cite{Mertsch:2018bqd}. 

Given the success of machine learning (ML) techniques in physical sciences, the question arises, how the modelling of the joint distribution of the CRE spectrum can benefit from state-of-the-art ML algorithms. 
Estimating the continuous density from a set of samples drawn from this distribution constitutes a kernel density estimation task. 
Conventional techniques often suffer from ``the curse of dimensionality'': the parameter volume grows exponentially with the number of parameter dimensions. 
In this work, we have trained a neural network as a probabilistic emulator to not only model the probability distribution of intensities at one energy, but also the correct correlations between energy bins. 
Our approach is based on factorising the joint distribution $p(\phi_1, \mathellipsis \phi_n)$ of intensities $\phi_i$ at $n$ energies $E_i$ into conditional probabilities $p(\phi_i | \phi_{i-1}, \mathellipsis \phi_1)$
\begin{equation}
p(\phi_1, \mathellipsis \phi_n) = p(\phi_1 | \phi_2, \mathellipsis \phi_n) \cdot p(\phi_2 | \phi_3, \mathellipsis \phi_n) \cdot \mathellipsis \cdot p(\phi_{n-1} | \phi_n) \, . 
\label{eqn:factorisation_joint_distribution}
\end{equation}
A particular network structure that allows to satisfy the autoregressive nature of eq.~\eqref{eqn:factorisation_joint_distribution} is Masked Autoencoder for Distribution Estimation (MADE)~\cite{2015arXiv150203509G}. 
While the output of the MADE as originally suggested consists of real numbers, representing, e.g. an estimator between 0 and 1, we employ Gaussian mixtures to model the conditional probabilities. 
Once the MADE has been trained to the results of MC simulations, the evaluation of the joint distribution is very fast (evaluating the likelihood of $10^6$ spectra takes only $\sim 5\,\unit{s}$ on a standard GPU). 
In addition to evaluating the joint distribution, we can also efficiently sample from it (sampling $10^6$ spectra takes only $\sim 10 \,\unit{s}$ on a standard GPU). 

A MADE trained on MC simulations for one fixed combination of CR parameters of course only represents the joint distribution of this parameter combination. 
The conditional nature of the MADE can however also be used to take into account the CR parameters as additional inputs. 
We have thus trained an extended MADE to a large set of MC simulations for a combination of CR parameters. 
This extended MADE then allows sampling from the joint distribution for arbitrary parameter combinations, efficiently interpolating between the parameter points that it was trained on. 
We have made our model, dubbed SECRET (Stochasticity Emulator for Cosmic Ray Electrons), available to the community for efficient stochastic modelling of the CRE spectrum. 

The remainder of the paper is structured as follows: 
In Sec.~\ref{sec:simulations} we describe our CRE simulations and discuss the parameter ranges that we considered. 
The MADE is introduced in Sec.~\ref{sec:MADE}, both in the conventional version that applies to one parameter combination and the extended version that allows for interpolations. 
Throughout, we quantify the accuracy of the method. 
We also provide a worked example, showcasing the use of the SECRET code. 
We conclude in Sec.~\ref{sec:conclusions}.

\section{Simulations}
\label{sec:simulations}

At high energies, the spectrum of cosmic ray electrons (CRE) is dominated by the nearest and youngest CR sources~\cite{1964ocr..book.....G}. These sources are effectively pointlike with a discrete distribution through our galaxy. Modelling the resulting intensity thus depends on the exact locations and ages of the contributing sources. This implies that the CRE intensity observable at earth consists of a broad featureless spectrum from old and far-away sources, as well as small-scale structures coming from the superposition of individual contributions. As the exact distribution of sources in space and time is not known, stochastic modelling becomes relevant. 

In this section we present our approach to modelling the CRE spectra from individual sources and the full intensity we can observe at earth, and discuss the relevant parametrisation that describes the resulting stochastic intensity.

\subsection{Cosmic Ray Transport}
\label{sec:transport}

The propagation of high-energy electrons is governed by the transport equation

\begin{equation}
	\frac{\partial n}{\partial t}
	- \nabla \cdot (\kappa \cdot \nabla n)
	+ \frac{\partial}{\partial E} (b(E) \, n) = q \,
\label{eq:Propagation}
\end{equation}
with the differential CR electron density $n = \frac{dN}{dE}$, the diffusion coefficient $\kappa$, energy loss term $b(E) = \frac{dE}{dt} < 0$, and a source term $q$\footnote{In the following we make the simplifying assumptions of omitting secondary production of $e^\pm$ during the propagation, as well as reacceleration, convection, and the effect from solar modulation. This treatment is justifiable at the highest energies, e.g.~\cite{Mertsch:2018bqd}, which we consider in this work.}.

We consider diffusion in a cylindrical halo of half-height $z_\mathrm{max}$ and radius $s_\mathrm{max}$.
In general, CR diffusion is anisotropic in the presence of a regular magnetic field. Here, we assume the diffusion coefficient to be scalar, which corresponds to considering only isotropic diffusion. Furthermore, we assume that the diffusion coefficient is sufficiently homogeneous on the scales we consider, and we neglect its spatial variations, which lets us simplify $\nabla \cdot (\kappa \cdot \nabla n) = \kappa \Delta n$.

The isotropic diffusion coefficient can be modelled by

\begin{equation}
	\kappa(E) = \kappa_0 \cdot \left( \frac{E}{1\unit{GeV}} \right)^\delta \, ,
\end{equation}
with normalisation $\kappa_0$ and spectral index $\delta$.

For simplicity, we assume that all sources have the same power-law energy spectrum with an exponential cutoff

\begin{equation}
	Q(E) = Q_*(E/E_*)^{-\gamma} \exp[-E/E_{\mathrm{cut}}] \, , 
\label{eq:SourceSpectrumPowerLaw}
\end{equation}
with a normalisation $Q_*$, a spectral index $\gamma$ and a cutoff energy $E_{\mathrm{cut}}$. We can factorise the source term $q$ into a source density $\sigma$ and the spectrum of each source: $q(\textbf{r},t,E) = \sigma (\textbf{r},t) \cdot Q(E)$.
For definiteness, we will below assume for the source spectrum, rate and spatial distribution to follow those of SNRs. However, we stress that our methodology applies to other potential classes of high-energy electrons and positrons, e.g.\ pulsar wind nebulae. 

In the following derivation of an analytical solution to eq.~\eqref{eq:Propagation}, we closely follow {Ref.}~\cite{Mertsch:2018bqd}. 
The energy loss term for high-energy relativistic electrons in the Klein-Nishina regime is given by

\begin{equation}
	b(E) = \frac{dE}{dt} = -\frac{4}{3} \sigma_T c \Gamma^2 \sum_{r \in \mathrm{ISRF}} U_r \left( 1 - \frac{63}{10}\frac{\Gamma\left< \epsilon_r^2 \right>}{m_ec^2 \left< \epsilon_r \right>} \right)\, ,
\end{equation}
where $\sigma_T$ denotes the Thomson cross-section, $c$ the speed of light, $m_e$ the electron mass, and $\Gamma = \frac{1}{1-\beta^2}$ is the relativistic Lorentz factor of the electrons. $U_r$, $\left< \epsilon_r \right>$ and $\left< \epsilon_r^2 \right>$ are the energy density, mean photon energy, and mean squared photon energy of the interstellar radiation field (ISRF) components, which we take from \cite{Mertsch:2018bqd}.

With these assumptions, we can find a solution to the propagation equation via the Green's function of eq.~\ref{eq:Propagation}.

\begin{equation}
	G_\mathrm{free}(\textbf{r} - \textbf{r}_0,t-t_0;E,E_0)
	= (4\pi l^2)^{-3/2} \frac{1}{|b(E)|}
	\exp\left[ -\frac{(\textbf{r}-\textbf{r}_0)^2}{4l^2}\right]
	\delta (t-t_0-\tau)
\label{eq:GreensFunctionFree}
\end{equation}
where we defined the diffusion loss length $l$ and energy loss time $\tau$ as

\begin{equation} 
	l^2 = l^2(E,E_0) \equiv \int_{E_0}^{E} dE' \frac{\kappa (E')}{b(E')} ;
	\quad \quad
	\tau = \tau (E,E_0) \equiv \int_{E_0}^{E} \frac{dE'}{b(E')} .
\label{eq:DefinitionsL^2Tau}
\end{equation}

We ensure the boundary conditions of $G(\pm z_{max}) = 0$ are satisfied via the method of mirror charges:

\begin{equation}
G(\textbf{r} - \textbf{r}_0,t-t_0;E,E_0) = \sum_{i=-\infty}^{\infty} (-1)^i G_{free}(\textbf{r} - [x_0,y_0,2iz_\mathrm{max}+(-1)^iz_0]^T,t-t_0;E,E_0)
\end{equation}

Finally, we obtain the spectrum from a single source $i$ at distance $s_i$ and age $t_i$:

\begin{align}
	n_i =& \int_{E}^{\infty} dE'_0 G(\textbf{r}-\textbf{r}_i,t-t_i;E,E'_0) Q(E'_0) \\
	=& \; (4\pi l^2)^{-1} e^{-s_i^2/(4l_i^2)}
	\frac{b(E_0)}{|b(E)|} Q(E_0) \left( (4\pi l_i^2)^{-1/2} \sum_{n=-\infty}^{\infty} (-1)^n
	e^{-(z-z_{i,n})^2/(4l^2)} \right) \\
	\equiv & n(s_i,t_i,E)
\label{eq:SingleSourceSpectrum}
\end{align}

With this functional form\footnote{For reference, individual spectra are displayed in figure 1 of~\cite{Mertsch:2018bqd} for different source distances and ages, following this functional form.}, we can confirm that the closest and youngest sources contribute at the highest energies and note that the maximum energy of a source is given by a sharp cutoff, with an approximate relation of $E_\mathrm{max}\propto\frac{1}{t}$.

Finally, we construct a stochastic Monte Carlo model of the all-electron spectrum at earth.
We simulate a large number of sources and add up their contributions to the total intensity.
The simulations of an ensemble of sources involve the following ingredients:

\paragraph{Ages}
We draw the source ages from a uniform distribution $t \in [0,t_\mathrm{max}]$. $t_\mathrm{max}$ is the maximum age of sources in the simulation.
The maximum energy at which a source can contribute is given by $E_\mathrm{max}(t) \approx (b_*t)^{-1}$ in the Thomson limit of $\frac{dE}{dt} = b_*E^2$.
Consequently, the minimum energy of interest defines the required minimum value of $t_\mathrm{{max}}$, as this parameter establishes the oldest sources capable of contributing at or above the given energy threshold.
It also determines the required number of sources in the simulation, following the relation $N_\mathrm{src} = \mathcal{R}_\mathrm{SN} \cdot t_\mathrm{max}$, where $\mathcal{R}_\mathrm{SN}$ denotes the supernova (SN) rate. To minimize the computational effort, we choose $t_\mathrm{max}$ as low as possible, while still ensuring completeness on the given energy range via the relation we just described.
For the following analysis, we require a minimum energy of $10^{1.5}\,\mathrm{GeV}$, resulting in $t_\mathrm{max} = 7\;\mathrm{Myr}$.

\begin{figure}[h] 
	\begin{subfigure}{0.5\textwidth}
		\centering
		\includegraphics[width=0.9\textwidth]{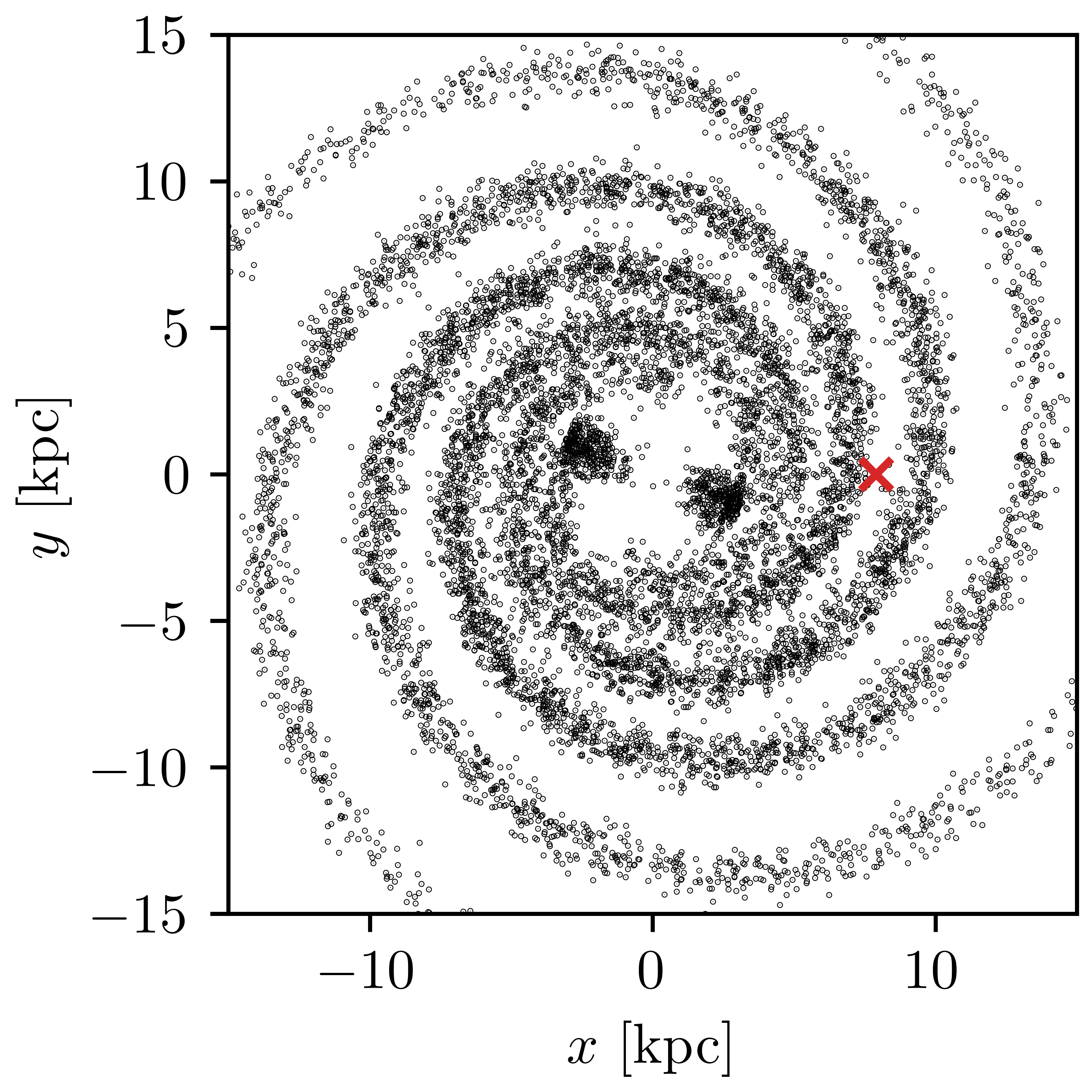}
		\phantomcaption{}
		\label{fig:SourceDistributionSpiral}
	\end{subfigure}
	\begin{subfigure}{0.5\textwidth}
		\centering
		\includegraphics[width=0.9\textwidth]{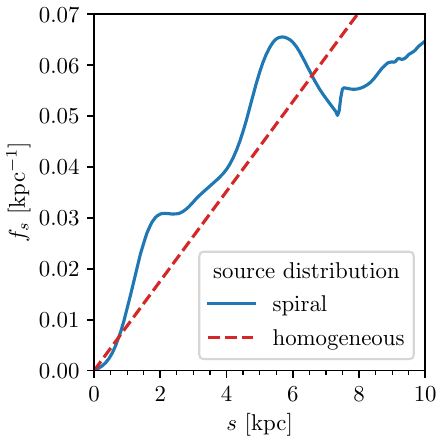}
		\phantomcaption{}
		\label{fig:SourceDistributionRadial}
	\end{subfigure}
	\caption{The spiral source distribution model of Ahlers et al. (2009)~\cite{Ahlers09}. \textit{Left:} The 2D milky way model. The distribution of samples follows the galactic arms and bar. The red cross marks the solar position at a distance of 7.9\unit{kpc} from the galactic center. \textit{Right:} Radial distribution of sources with respect to the position of the solar system, i.e. the distribution of source distances to us. The blue line is obtained by integrating the spiral 2D distribution over the polar angle around the solar position. For comparison, the red dashed line shows the radial distribution for a homogeneous source distribution in the galactic plane.}
	\label{fig:SourceDistribution}
\end{figure}

\paragraph{Distances}
For the spatial source distribution, we adopt the model by \cite{Ahlers09}. It consists of four logarithmic spiral arms \cite{Vallee05} and a radial dependence of $f(r) = A \sin \left( \frac{\pi r}{r_0} + \theta_0 \right) e^{-\beta r}$, where $r$ is the galactocentric radius and the parameters take the values $A = \qty{1.96}{kpc^{-2}}$, $r_0 = \qty{17.2}{kpc}$, and $\theta_0 = 0.08$. The 2D distribution is shown in Fig.~\ref{fig:SourceDistributionSpiral}.
Since we are only interested in the radial distances to Earth, we integrate over the angle, resulting in the distribution $f_s$ in Fig.~\ref{fig:SourceDistributionRadial}. We only simulate nearby sources up to a maximum distance $s_\mathrm{max} = 10\unit{kpc}$, which reduces the number of simulated sources by a factor of $\int_{0\unit{kpc}}^{10\unit{kpc}} \mathrm{d}s f_s = 0.44$.

\paragraph{Causal cut}
The transport equation allows for solutions that violate causality, i.e. non-zero solutions from outside the light cone. To prevent this, we manually remove sources for which the relation $c\cdot t \leq s$ does not hold. This typically affects only $\sim0.4\%$ of sources in our setup.

\paragraph{Total intensity}
Finally, we obtain the total intensity by adding all contributions from $N_\mathrm{src}$ together and multiplying with a constant flux factor:

\begin{equation}
	\phi (E) = \frac{c}{4\pi} \sum_{i=1}^{N_\mathrm{src}} n (s_i,t_i,E)
\label{eq:SummedFlux}
\end{equation}
where $n (s_i,t_i,E)$ is the Green's function from eq. \ref{eq:SingleSourceSpectrum}.

\begin{figure}[h]
	\centering
	\includegraphics[width=\textwidth]{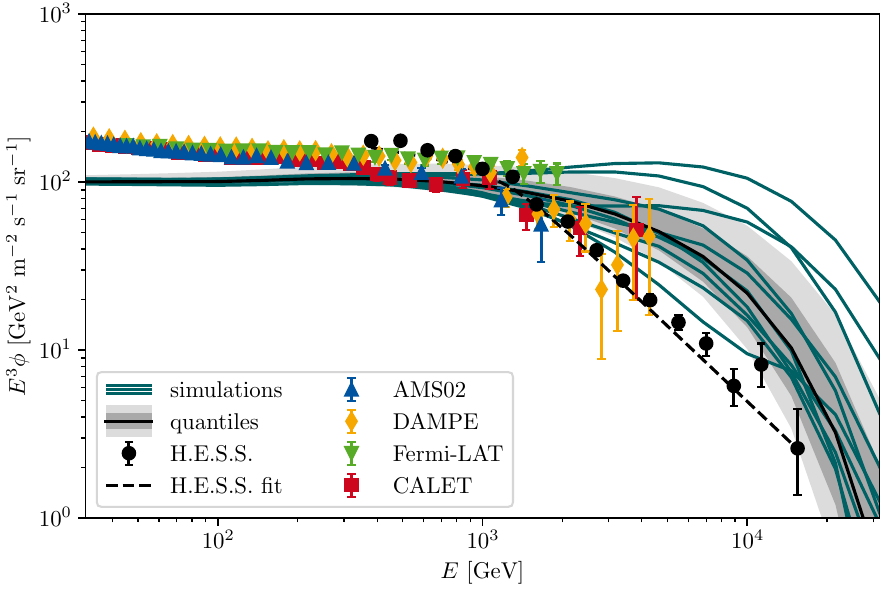}
	\caption{An ensemble of simulated spectra, along with measurements from different experimental collaborations obtained from the Cosmic-Ray Data Base \citep{CRDB4.0}, plus the H.E.S.S. measurement and fit~\cite{2024PhRvL.133v1001A}. The respective error bars are the squared sums of statistical and systematic uncertainties. The black line is the median of the distribution, and the grey areas show the 68~\% and 90~\% quantile bands. The fit of the ensemble to the datapoints could be improved by choosing different physical parameters, see section~\ref{sec:parameters}.
    The distribution is very narrow at low energies, where a lot of sources contribute and the contribution of individual sources average out. At \unit{TeV} energies, there is a large spread of possible intensities and significant variation between realisations, as single nearby sources start to dominate and stochastic effects become important.}
\label{fig:SimulatedEnsemble}
\end{figure}

The result of such a simulation is shown in figure~\ref{fig:SimulatedEnsemble}. The lines show examples from the Monte Carlo-generated dataset. Because the distributions are non-Gaussian and the variance is divergent~\cite{Mertsch:2010fn}, we instead quantify the spread and centre of the distribution by quantiles. The black line denotes the median, which deviates slightly from the mean. The grey bands denote the 68~\% (90~\%) bands centered around the median.
\footnote{Note that the physical parameters chosen in this setup best match the H.E.S.S. measurement~\cite{Mertsch:2018bqd} and the disagreement with other experiments at low energies can be reduced with different physical assumptions, see section~\ref{sec:parameters}.}

\subsection{Model Parameters}
\label{sec:parameters}

We are aiming for an emulator that covers as much of the space of physical models as possible. Therefore, we want our stochastic model parametrisation to allow for sufficient freedom. For this we choose to vary five parameters in our simulations: the SN rate $\mathcal{R}_\mathrm{SN}$, the source spectrum cutoff energy $E_\mathrm{cut}$ and spectral index $\gamma$, and the diffusion coefficient spectral index $\delta$ and normalisation $\kappa_0$. 

To get even phase-space coverage over the entire five-dimensional space, we simulate an ensemble of realisations on a hypercubic grid, such that every possible combination is contained in the dataset. As a result we obtain a binned set of intensity ensembles.
Note that the full hypercubic setup also includes combinations of parameters that are incompatible with data. For example, low values of $\gamma$ combined with the lowest values of $\delta$ produce a significantly harder spectrum than the $\sim E^3$ scaling observed in measurements. This also entails some unlucky parameter combinations, especially towards the corners of the phase space, that are extreme in their shape and structure.

\begin{table}[htbp]
	\centering
	\begin{tabular}{ccc}
		\hline
		Parameter & Range & Step size \\ \hline
		$\mathcal{R}_{\mathrm{SN}} [10^4\unit{Myr^{-1}}]$ & 0.5--3.0 & 0.5 \\
		$\log_{10}(E_\mathrm{cut} [\unit{GeV}])$ & 3.5--5.5 & 0.25 \\
		$\gamma$ & 1.8--2.6 & 0.2 \\
		$\delta$ & 0.2--1.0 & 0.2 \\
		$\kappa_0 [10^{28}\unit{cm^2s^{-1}}]$ & 1.0--9.0 & 2 \\ \hline
	\end{tabular}
    \caption{Simulation parameters and their respective ranges and binning. Their physical properties are outlined in section~\ref{sec:transport}.}
	\label{tab:ParamRanges}
\end{table}

\begin{figure}[h]
	\begin{subfigure}[t]{0.5\textwidth}
		\centering
		\includegraphics[width=0.9\textwidth]{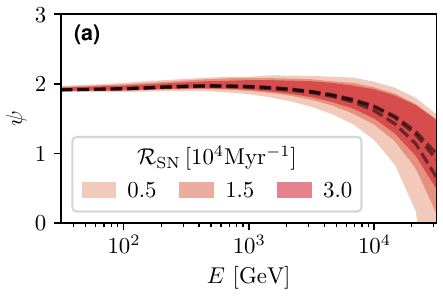}
		\phantomcaption{}
		\label{fig:SpectraVaryingSNrate}
	\end{subfigure}
	\hfill
	\begin{subfigure}[t]{0.5\textwidth}
		\centering
		\includegraphics[width=0.9\textwidth]{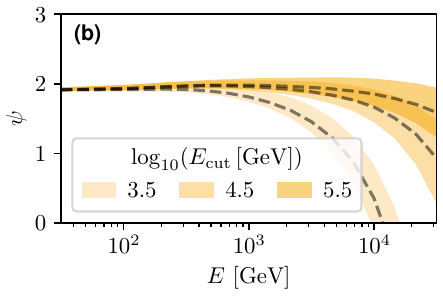}
		\phantomcaption{}
		\label{fig:SpectraVaryingEcut}
	\end{subfigure}

    \begin{subfigure}[t]{0.5\textwidth}
		\centering
		\includegraphics[width=0.9\textwidth]{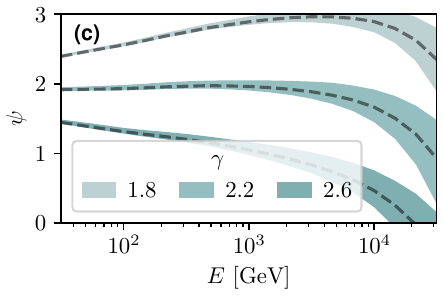}
		\phantomcaption{}
		\label{fig:SpectraVaryingGamma}
	\end{subfigure}
	\hfill
    \begin{subfigure}[t]{0.5\textwidth}
		\centering
		\includegraphics[width=0.9\textwidth]{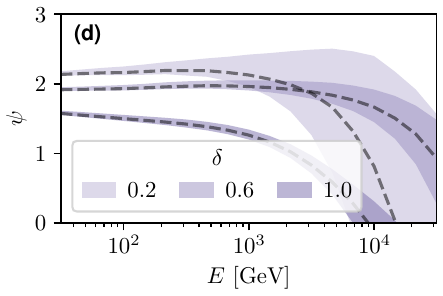}
		\phantomcaption{}
		\label{fig:SpectraVaryingDelta}
	\end{subfigure}

	\centering
    \begin{subfigure}[t]{0.5\textwidth}
		\centering
		\includegraphics[width=0.9\textwidth]{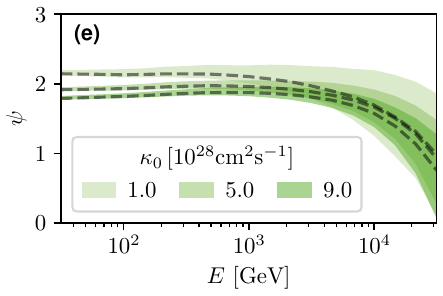}
		\phantomcaption{}
		\label{fig:SpectraVaryingD0}
	\end{subfigure}
	
	\vspace{-0.5cm}
	\caption{Variation of stochastic spectra with individual parameters, with the log of the rescaled flux on the y-axis, defined by $\psi(E) \equiv \log_{10}\left(\frac{E^3\cdot \phi(E)} {\unit{GeV^2m^{-2}s^{-1}sr^{-1}}}\right)$. Each plot shows the 90\% bands (coloured bands) and medians (dashed line) of an ensemble of realisations for different values of a given parameter. All remaining parameters are kept constant at $\mathcal{R}_{\mathrm{SN}}=2\cdot10^4\unit{Myr^{-1}}$, $E_\mathrm{cut}=10^{4.5}\unit{GeV}$, $\gamma=2.2$, $\delta=0.6$, and $\kappa_0=5\cdot10^{28}\unit{cm^2s^{-1}}$.}
	\label{fig:SpectraVaryingParameters}
\end{figure}

The adopted ranges and binning of each parameter are detailed in table~\ref{tab:ParamRanges} and motivated in the following.
The galactic SN rate $\mathcal{R}_{\mathrm{SN}}$ is estimated to be about $(2\cdot 10^4 - 3\cdot10^4)\unit{Myr^{-1}}$ \cite{GalacticSNRate, 2021NewA...8301498R}, where the exact value measured depends on the method used (see Ref.\cite{2021NewA...8301498R} for a discussion). From the perspective of stochasticity, we have a motivation to consider lower values \cite{Mertsch:2018bqd}, down to $0.5\cdot10^4\unit{Myr^{-1}}$. 
For an upper limit we set $\mathcal{R}_{\mathrm{SN}} = 3\cdot10^4\unit{Myr^{-1}}$, which is in agreement with the measurement and simultaneously computationally feasible, as the relation $N_\mathrm{src} = \mathcal{R}_{\mathrm{SN}} t_\mathrm{max}$ leads to an increasing computational effort with increasing $\mathcal{R}_{\mathrm{SN}}$. 
A theoretical motivation for the maximum energy of particles accelerated in SNR of $(10^{4} - 10^{5})\unit{GeV}$ can be derived as shown in \cite{Lagage83}. We select the source spectrum cutoff energy $E_\mathrm{cut}$ from $(10^{3.5}-10^{5.5})\unit{GeV}$ with a logarithmic binning. 
There currently are no strong constraints for the source spectral index $\gamma$ for electrons. State-of-the-art models \cite{Schwefer:2022zly} give a best-fit value of $2.5$, while our simulations can roughly reproduce the observed energy scaling down to about $2$. We allow for $\gamma$ to lie in the range $1.8 - 2.6$.
Finally, the diffusion coefficient parameters $\delta$ and $\kappa_0$ were chosen to generously cover the found values by state-of-the-art models \cite{Schwefer:2022zly} of $\sim 0.2 - 0.5$ and $5\cdot10^{28}\unit{cm^2s^{-1}}$ respectively. As will become apparent later in the paper (cf. figure \ref{fig:SpectraVaryingParameters}), low values of $\delta$ are unrealistic, so we extend the range only upwards. We chose $\delta \in [0.2, 1.0]$ and $\kappa_0 \in [1, 9] \cdot10^{28}~\unit{cm^2s^{-1}}$.
Regarding the binning of each parameter, the limiting factor is the dataset size. The number of phase space points is given by the product of all bin numbers in each dimension, so increasing the resolution would rapidly increase the dataset size. We decided on the bin sizes listed in Tab.~\ref{tab:ParamRanges}, results in a total of 6750 parameter combinations.

The effect each of the parameters individually has on the overall intensity distribution is shown in  figure~\ref{fig:SpectraVaryingParameters}.
As expected, the SN rate mainly changes the level of stochastic effects, i.e. the width of the distribution at high energies\footnote{The effect on the normalisation has been factored out.}.
The effect of the source spectrum parameters is also intuitively understood: The source spectral index is mostly reflected in the spectral index of the final spectrum (and by extension also the normalisation in log space), and the cutoff energy determines the position of the cutoff.

For the diffusion coefficient parameters we can observe purely stochastic effects: A small diffusion coefficient drastically reduces the radius in which sources can contribute at the highest energies, which in turn reduces the number of sources. This explains the premature cutoff and widening of the distribution for small $\delta$ and $\kappa_0$. This is a feature that is not present in the prediction of the expectation value for which an analytical approximation is possible~\cite{Mertsch:2010fn}.

Finally, the energy range is fixed to a logarithmic binning between $(10^{1.5}-10^{4.5})\unit{GeV}$ according to the following considerations: On the lower end, the lowest energy bin determines the maximum source age $t_\mathrm{max}$, and thereby the number of sources we need to simulate. Thus, extending the range to lower energies can drastically increase the computational effort required.
Secondly, at the highest energies, the high energy cutoff can lead to numerical issues, specifically for the standardisation of the data for the resolution of the neural network input. As we are most interested in the cutoff region where stochastic effects are strongest, the effects on the intensity beyond the cutoff is less important and can be considered negligible for practical purposes in this work. Intensities below $\psi\sim \mathcal{O}(1)\unit{GeV^2cm^2s^{-1}sr^{-1}}$ at tens of \unit{TeV} are far below the sensitivity of current and even upcoming experiments for the foreseeable future. Finally, the difficulty of a density estimation task increases exponentially with the dimensionality of the problem. Therefore, we limit the energy resolution to 19 bins.

\section{MADE}
\label{sec:MADE}

Having established the inherent stochastic properties of CR electron spectra and the MC-approach we use for simulations, we now want to build a model for the probability distribution of intensities.
We choose a machine learning approach due to the effectiveness in dealing with high-dimensional data and high flexibility and expressivity of neural networks. Namely we use the Masked Autoencoder for Distribution Estimation (MADE) \cite{2015arXiv150203509G} to perform a density estimation task. This model can learn and calculate efficiently the likelihood of datapoints and can generate samples from the learned distribution.
This will alleviate the necessity to rerun the computationally expensive MC simulations every time one is interested in a slightly different physical model with different physical parameters.

In this Section, we first demonstrate MADE's effectiveness on this task by accurately modeling the distribution of spectra on a training set of simulated intensities and evaluating its performance in detail. We then extend the method to a much more flexible model that simultaneously learns the distribution over many different physical models by including varying physical parameters in the dataset and conditioning on them during training. The resulting network is able to flexibly predict likelihoods and generate samples conditioned on the underlying parameters.

\subsection{Method}

In this Section we introduce the MADE and discuss its architecture and how it models arbitrary multivariate densities. Furthermore we extend the original MADE architecture to real-valued data by utilising techniques from mixture density networks. While we describe and explain our model, we will often refer to standard ML techniques and assume the reader has some familiarity with basic ML concepts. For an introduction to ML see \cite{Bishop+Bishop} or \cite{Goodfellow-et-al-2016}.

Every multivariate probability can be factorised into conditional probabilities using the chain rule:

\begin{equation} 
    p(\vec{x}) = p(x_2,\dots ,x_D | x_1) \cdot p(x_1) = p(\vec{x}_{>1} | x_1) \cdot p(x_1) = \prod_{i=1}^{D} p(x_i | \vec{x}_{<i}) \, ,
\label{eq:ConditionalFactorisation}
\end{equation}
where $\vec{x}_{>i} \equiv \{ x_{i+1}, x_{i+2}, \mathellipsis x_D \}$ and $\vec{x}_{<i}\equiv \{ x_1, x_2, \mathellipsis x_{i-1} \}$ denote subsets of the elements of a $D$-dimensional vector $\vec{x}$.
In this factorisation, the ordering of variables is arbitrary, e.g. $p(x_1, x_2) = p(x_1 | x_2) p(x_2) = p(x_2 | x_1) p(x_1)$.
Bringing the probability into this form allows one to easily and exactly marginalise over (and condition on) given variables, if an ordering has been chosen such that the respective variables are last (first) in the ordering,
%
\begin{align}
    p(\vec{x}_{\leq j}) &=
    \prod_{i=1}^{j} p(x_i|\vec{x}_{<i}) \, , \\
    p(\vec{x}_{\geq j} | \vec{x}_{<j}) &=
    p(\vec{x}_{<j}) \prod_{i=j}^{D} p(x_i|\vec{x}_{<i}) \, .
\label{eq:ProbabilityConditioningMarginalisation}
\end{align}

In this work, we use the Masked Autoencoder for Distribution Estimation (MADE) \cite{2015arXiv150203509G}.
It builds on a basic fully connected feed-forward neural network (NN) also known as Multilayer Perceptron (MLP). We can represent a MLP as a graph structure with layers of nodes and connections that represent the network weights. In MADE, the connections between nodes are masked in such a way that the outputs are autoregressive, i.e. each output only depends on previous outputs.
We use this to model each conditional probability of the autoregressive factorisation of eq. \ref{eq:ConditionalFactorisation}, where every output node corresponds to the (conditional) probability of one input variable.
The network is trained by minimising the negative log-likelihood $-\log p(\vec{x}|\theta)$ of the dataset under the model, which is characterised by its parameters $\theta$.

\begin{figure}[h]
	\centering
	\includegraphics[width=\columnwidth]{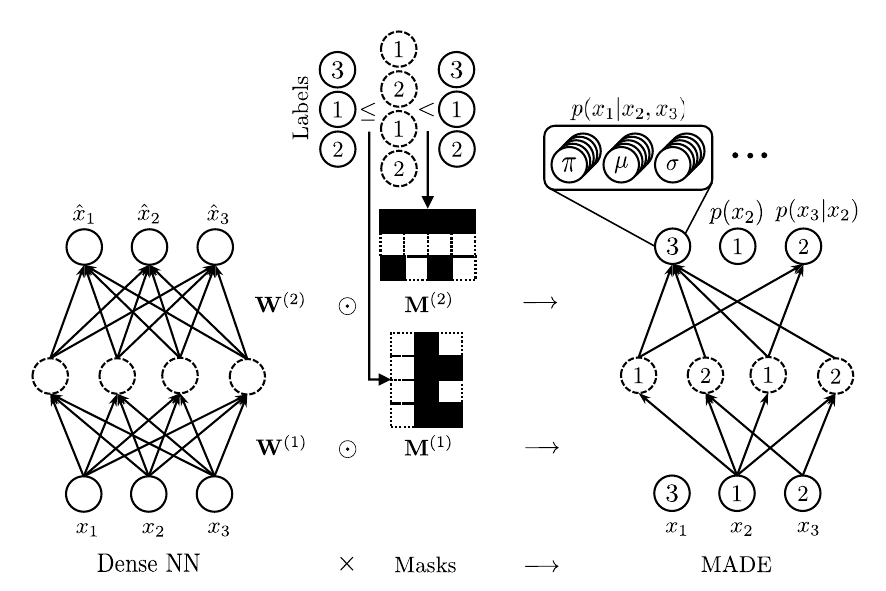}
	\caption{A depiction of the construction of the MADE architecture, adapted from \cite{2015arXiv150203509G}. \textit{Left:} A standard fully connected neural network with weight matrices $\textbf{W}^{(i)}$ and output nodes that correspond to the inputs. \textit{Middle:} Integer labels are assigned to each node, with a random permutation for input and equivalently output layer, and uniformly sampled for the hidden layer. Binary masks $\textbf{M}^{(i)}$ are generated according to $\leq$ and $<$ relations between the labels. \textit{Right:} The weights of the NN are multiplied element-wise $\textbf{W}^{(i)} \odot \textbf{M}^{(i)}$, effectively masking out certain weights and severing connections between nodes. The resulting network obeys the information flow dictated by the labels and thus fulfills the autoregressive property. Furthermore, each output node is replaced by a set of nodes representing the parameters of a mixture distribution.}
\label{fig:MADEArchitecture}
\end{figure}

To obtain valid autoregressive outputs, MADE utilises a specific masking on its weights. The procedure is visualised in Fig. \ref{fig:MADEArchitecture} and described in the following.

Firstly, all nodes are labeled with integer values $m$. The labels of the input layer are a random permutation of $\{1,2,\dots,D\}$. This is what we call the ordering of variables, as these labels will determine the ordering in the autoregressive factorisation and the position each input will have in that order. The output nodes get the same $m$ as the input, as each one corresponds to its respective input node.
The labels for the hidden nodes are sampled randomly from a uniform distribution between $1$ and $D$ ($\mathcal{U}_{\mathbb{N}}(1, D)$).

Then, masks are constructed from these labels, according to the conditions

\begin{align} 
    M^{(1)}_{ij} = \begin{cases}
        1 & \text{for\ } m^{(0)}_i \leq m^{(1)}_j\\
        0 & \text{otherwise}
    \end{cases}
    \qquad\text{and}\qquad
    M^{(2)}_{ij} = \begin{cases}
        1 & \text{for\ } m^{(1)}_i < m^{(2)}_j\\
        0 & \text{otherwise}
    \end{cases}.
\label{eq:MADEMasks}
\end{align}

Masks are applied to the NN weights $W_{ij}$ during the forward pass: $W'^{(1)}_{ij} = M^{(1)}_{ij} \odot W^{(1)}_{ij}$ and $W'^{(2)}_{ij} = M^{(2)}_{ij} \odot W^{(2)}_{ij}$, where $\odot$ denotes the element-wise product.
The masking effectively removes connections between nodes, limiting the flow of information through the network. The conditions ensure that nodes with label $m$ only feed into nodes with a higher or equal number.
For the last layer, the mask is created with the alternative, stricter condition $m^{(1)} < m^{(2)}$, such that there are no direct connections from one input node at position $m$ in the ordering to its corresponding output node at position $m$. Creating masks in this way ensures that only inputs with label $<d$ can transfer information into output $p(x_d|\vec{x}_{<d})$, which is exactly the restriction requested by the autoregressive conditional.

For example, the first output $p(x_1)$ does not have any connections to any nodes, because no previous information is available. Similarly, the last input $x_D$ has no connection to the hidden layers, as its information never enters the output probabilities. Also, the first input $x_1$ enters every output except the first one.
This is illustrated in Fig.~\ref{fig:MADEArchitecture}.

Following this procedure ensures that the autoregressive property is fulfilled, making the output a valid probability. Furthermore, this type of masking can be easily generalised to deeper networks, where the masks between hidden layers are simply constructed from the same condition as in the first layer. The output can be computed with only a single forward pass, making it relatively efficient to train and evaluate. Sampling from the network has to be done sequentially and requires $D$ forward passes. For each output, the predicted distribution is calculated by passing the input through the network. From this distribution, a value for $x_i$ is drawn, which is then fed again into the NN to draw the value of $x_{i+1}$. This is repeated until the full vector is sampled.

\subsubsection{Real-valued Outputs}

In its original form, MADE only supports binary outputs. Each output distribution is defined by one node, which is then interpreted as a binary probability. However, our physical data is not binary, but real-valued.
We use a method similar to \cite{RNADE}, where the outputs are instead interpreted as the parameters of some parametrised distribution.

In our case, there is no analytical expression for the shape of the conditional probabilities, so we want the model to be able to approximate arbitrary distributions. For that, we chose a mixture of Gaussians (MoG).

\begin{equation}
    p(x_d|\vec{x}_{<d}) = \sum_{k=1}^{K} 
    \pi_{d,k} \, \mathcal{N}(x_d;\mu_{d,k},\sigma^2_{d,k}),
\label{eq:GaussianMixture}
\end{equation}
which is a superposition of $K$ Gaussians with three parameters each: a relative amplitude $\pi_{d,k}$, a mean $\mu_{d,k}$, and a standard deviation $\sigma_{d,k}$.

The number of output nodes computes as three, the product of the number of parameters, times $K$, the number of mixtures, times $D$, the number of dimensions.
A one-hidden-layer MADE Gaussian-mixture model with $D$-dimensional input $\vec{x}$, $H$ hidden nodes, and $K$ mixture components is given by:

\begin{align} 
    h_j &= \mathrm{sigmoid} \left( \sum_{i=1}^{D} (\vec{W}^{(1)}_{ij} \odot \vec{M}^{(1)}_{i,j}) \cdot x_i + \vec{W}^{(1)}_{0j} \right) \\
    \pi_{d,k} &= \mathrm{softmax} \left( \sum_{j=1}^{H} (\vec{W}^{(\pi)}_{jd,k} \odot \vec{M}^{(2)}_{jd}) \cdot h_j + \vec{W}^{(\pi)}_{0d,k} \right) \\
    \mu_{d,k} &= \left( \sum_{j=1}^{H} (\vec{W}^{(\mu)}_{jd,k} \odot \vec{M}^{(2)}_{jd}) \cdot h_j + \vec{W}^{(\mu)}_{0d,k} \right) \\
    \sigma_{d,k} &= \exp \left( \sum_{j=1}^{H} (\vec{W}^{(\sigma)}_{jd,k} \odot \vec{M}^{(2)}_{jd}) \cdot h_j + \vec{W}^{(\sigma)}_{0d,k} \right)
\label{eq:MLOneLayerMADE}
\end{align}

where $\pi$ and $\sigma$ use a softmax and exponential activation respectively to ensure they meet the requirements of $\sum_{k=1}^K \pi_{d,k} = 1$ and $\sigma_{d,k} > 0$. The masks $\vec{M}^{(2)}$ are copied for all output nodes belonging to one conditional probability.

\subsection{Single Point MADE}

First, we verify the method by learning the probability distribution for a fixed set of parameters. We demonstrate the capabilities of MADE for modelling stochastic intensities in this setup and evaluate its accuracy.

\subsubsection{Dataset}

We adopt the parameters of \cite{Mertsch:2018bqd} given by a supernova rate of $\mathcal{R}_{\mathrm{SN}}=4.55\times10^4\unit{Myr^{-1}}$, source spectrum parameters $E_{\mathrm{cut}}=1\times10^4\unit{GeV}$ and $\gamma=2.2$, as well as parameters of the diffusion coefficient $\kappa_0=3\times10^{28}\unit{cm^2/s}$ and $\delta=0.6$.\footnote{Note that the supernova rate is larger than stated in the paper by a factor of 2.27. This is based on the normalisation of the source distribution, which was not properly taken into account.}
Our simulations contain $10^5$ individual realisations. With the optimised $t_\mathrm{max}=7\unit{Myr}$, every realisation contains $\sim3.2\times10^5$ individual sources.

We define the energy grid with logarithmic spacing between $10^{2.4}$-$10^{4.2}\,\unit{GeV}$ with 19 bins, by which we cover the same energy range as \cite{Mertsch:2018bqd}

Before training, we have to preprocess the inputs. Making sure that all values are on the same scale generally improves the training process. Especially when values span many orders of magnitude, it is advisable to preprocess the data in a suitable way.
For this reason, we do not input the intensities $\phi(E)$ directly into the network, but instead only consider the intensities in rescaled log space: 
\begin{equation}
    \psi(E) \equiv \log_{10}\left( \frac{E^3\cdot \phi(E)}{1 \unit{GeV^2m^{-2}s^{-1}sr^{-1}}}\right).
    \label{eq:dimless_flux}
\end{equation} Note that this is a common way to represent power-law spectra, equaling the quantity as plotted on the y-axis in Fig.~\ref{fig:SimulatedEnsemble}.

Additionally, we standardise all intensities individually for every energy bin by subtracting the mean $\mu(E)$ and dividing by the standard deviation $\sigma(E)$, both of which are calculated empirically from the dataset at every energy bin: $x(E) \equiv \frac{\psi(E)-\mu(E)}{\sigma(E)}$.
While the true standard deviation is undefined as the integral $\int \phi^2 p(\phi)$ diverges \cite{Mertsch:2010fn}, the empirical value is appropriate for this purpose.
Both these transformations are trivially invertible, which is required to obtain physical intensities from the network outputs.

To make sure that the model performs well on unseen data, we set 10\% of the dataset apart for validation during training. Additional detail on the implementation can be found in the accompanying \href{https://git.rwth-aachen.de/pmertsch/secret}{GitHub repository}.

We optimised the network architecture by changing the hidden layer size, number of hidden layers, number of masks, and number of output components, and evaluating its performance under a standardised classifier, which is detailed later in this Section. The resulting best-performing model has one hidden layer with 200 nodes, 10 output components, and utilises only one fixed mask.

Furthermore, we found it to be imperative to choose a random permutation ordering instead of the natural incremental ordering, for which performance was much worse. In general, we can assume that some masks work better in this specific application than others. Our experiments with linear ordering indicate that the ordering should not be too regular. However, we did not explore and optimise this further and kept a completely random permutation ordering without additional constraints, which led to satisfactory results.

\subsubsection{Results}

In this section, we evaluate the performance of MADE performance by comparing its samples to simulations. Sampling $10^4$ data points from the network takes only a few seconds on a GPU, which corresponds to a speedup of $\sim\mathcal{O}(10^4)$ compared to running the Monte Carlo code.
In the following, we show the compatibility of samples and simulations in various ways.

\begin{figure}[h]
	\centering
	\includegraphics[width=0.8\textwidth]{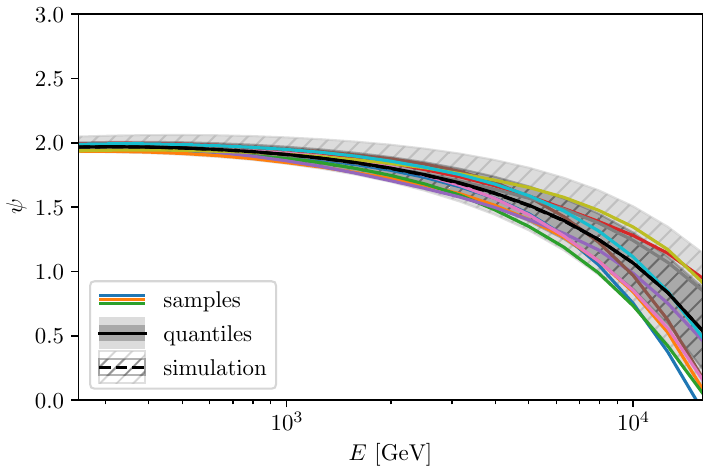}
	\caption{The distribution learned by MADE. The coloured lines show samples from MADE with the dimensionless flux defined by eq. \ref{eq:dimless_flux} on the y-axis. The black line (grey bands) represents the median ($68\%$ and $90\%$ quantiles) of MADE samples. The dashed line (hatched regions) indicate the median (same quantiles) of our simulations, which perfectly overlap with the learned distribution over the entire energy range.}
	\label{fig:MADEPredictionsSpectra}
\end{figure}

The ensemble of generated samples can be seen in Fig.~\ref{fig:MADEPredictionsSpectra}. The coloured lines show random example spectra and the grey bands are the 68\% and 90\% regions. The true median/quantile bands are indicated by the dashed line/hatched areas. The MADE samples overlap almost perfectly by eye with the simulations. The quantiles alone do not give us the full information about the distribution, and we have to also consider the full marginal distributions.

\begin{figure}[h]
	\centering
	\includegraphics[width=\textwidth]{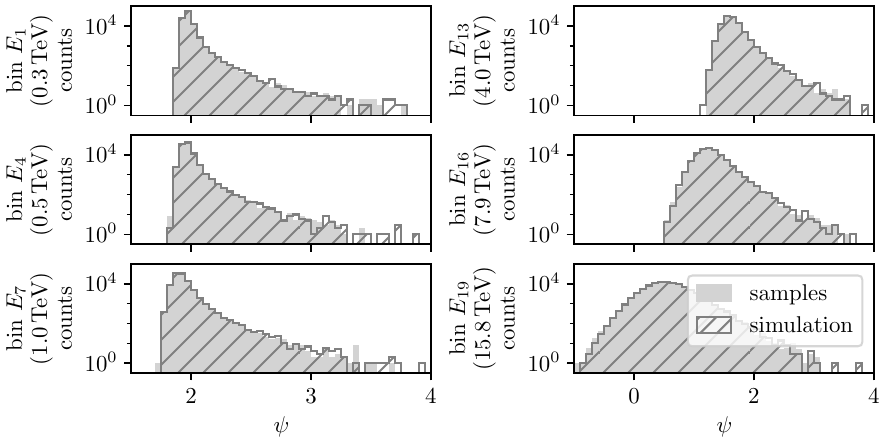}
	\caption{1D marginal dimensionless flux distributions of MADE samples (grey area) and simulations (hatched area) for some of the energy bins. The distribution is highly non-gaussian with heavy tails towards high values of $\psi$.
    The learned density overlaps exceedingly well with the target density for the bulk of the distribution. Slight deviations only appear in the tails with low statistics. Due to the inherent sparsity of datapoints no density estimation will be able to model the tails perfectly. However, MADE appears to be accurate up to the single counts regime.}
	\label{fig:MADEPredictions1DHists}
\end{figure}

Fig.~\ref{fig:MADEPredictions1DHists} shows the 1D marginalised histograms of intensities for some energy bins. The network samples (grey area) overlap very well with the simulations (hatched area). Slight differences can be seen towards the edge of the distributions, specifically in the long tails. This is because the number of training samples gets more and more sparse, and so the NN can no longer learn the density accurately. In the tail regions with very low sample density, the NN will never be able to model the density perfectly. In the final training shown here, we were able to cover the tails fairly well up to the highest intensity values, which agree up to Poisson uncertainties.
Since the 1D marginals do not show any of the correlations between energy bins, which are crucial in this case, we show the pairwise 2D marginals of both the original and learned distributions in the Appendix in Fig.~\ref{fig:MADEPredictionsCorner}. They agree very well and indicate that the network is indeed able to accurately reproduce the non-linear correlation structure of intensities at different energies.

\begin{figure}
    \centering
    \includegraphics[width=0.5\textwidth]{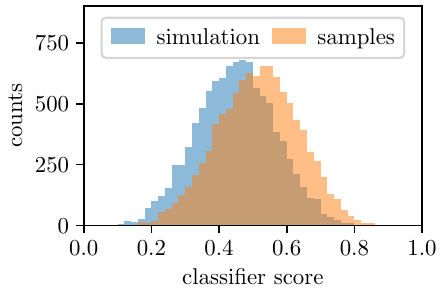}
    \caption{Classifier scores for both the simulated and MADE generated datasets. The score can be interpreted as the probability that the discriminator assigns to inputs for belonging to the MADE generated set. Both distributions are centered close to 0.5 and the strong overlap indicates that generated samples are hard to distinguish from the simulations.}
    \label{fig:ClassifierScores}
\end{figure}

Finally, we quantify the quality of samples via its robustness against a discriminator.
For that, we train a separate model to distinguish between simulated and MADE-generated samples. The accuracy of the discriminator is a way of quantifying the quality of MADE predictions that does not depend on extrinsic quantities: the lower the achieved accuracy, the harder it is to tell the two sets apart, the better the samples. This allows us to compare different architectures and adjust design choices, such as the number of layers. The best-performing model is relatively simple, with only one hidden layer. Increasing the number of layers or the width of each layer did not significantly improve the results.

The classifier is a basic two-layer neural network with a single output. The output is passed through a sigmoid function, such that the score $y\in [0,1]$ it assigns can be interpreted as the probability of belonging to one of the two classes. It is trained to minimise the cross-entropy loss $\mathcal{J}(y,t) = -\log( y^t \cdot (1-y)^{1-t}) = -t\cdot\log y - (1-t)\cdot\log(1-y)$ with true labels $t\in\{0,1\}$, which is a standard optimisation goal for classification tasks.

To ensure a fair comparison between architectures, the training was standardised for all NNs and trained for the same number of epochs, regardless of how training progressed in eaech run. This also means that the training is not necessarily fully converged, but the classification accuracy is still valid as the defining measure for relative sample quality.

The resulting classification score distribution for the best-performing architecture can be seen in Fig.~\ref{fig:ClassifierScores}, which corresponds to an accuracy of $62\%$. While this value has no inherent meaning or interpretation in terms of sample quality, it does serve as a metric for the purpose of ranking architectures relative to each other.
Note also that the discriminator is not optimal and its accuracy represents a lower bound on the achievable separability between the datasets. It would likely be possible to devise a more effective classifier with some optimisation.

\subsection{SECRET}
\label{sec:SECRET}

In the previous Section, we have shown how density estimation with MADE can be efficiently used to model stochastic CR electron spectra. However, while this approach allows for sampling random intensities accurately and many orders of magnitude faster than performing a full simulation, it also has one significant shortcoming: it does naturally not generalise beyond its training data. Changing the underlying physical model would require rerunning simulations and retraining the NN, which is computationally expensive and unfeasible on a large scale. Instead, we want a model that unites an ensemble of MADEs into one single model.

In this Section, we introduce SECRET, the Stochasticity Emulator for Cosmic Ray ElecTrons. It is a density estimator for stochastic CR electron intensities that can model a variety of physical scenarios by conditioning the network on the underlying parameters.
SECRET is intended to be used as a tool for quickly, efficiently, and flexibly emulating stochastic spectra, eliminating the need to set up and perform time-consuming simulations from the ground up.

The idea is to create a model that can produce stochastic spectrum realisations for different values of the underlying physical parameters of our simulations. We achieve this by devising an extended version of MADE that takes the parameters as additional inputs.
In principle, the split between predictive intensity dimensions and auxiliary parameter dimensions is arbitrary and the NN architecture is oblivious to the nature of its inputs. They are only separated by our physical interpretation and a different correlation structure.
Choosing an ordering where the CR intensity dimensions are conditioned on additional dimensions, the predicted spectra can be conditioned on additional parameters, effectively resulting in an ensemble of MADE models for different parameter combinations. The conditioning can be done exactly due to the nature of the autoregressive factorisation.
The network is trained on a large-scale dataset that contains stochastic realisations of spectra from a variety of different simulations, as well as the corresponding parameter values. We expect the network to be able to interpolate between the training data points, such that it learns the distribution of all points within the trained phase space volume.

\begin{figure}[h]
	\centering
	\includegraphics[width=\textwidth]{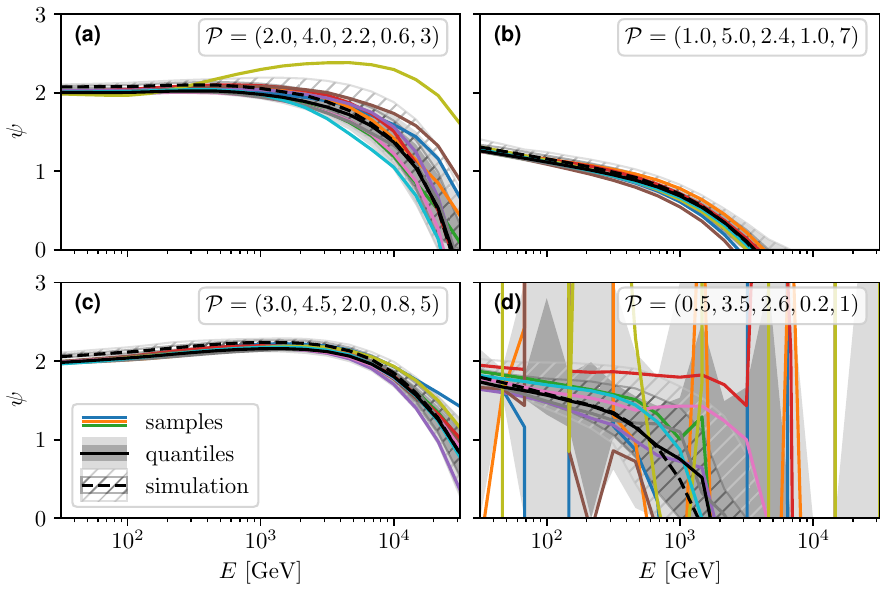}
	\caption{Some examples of SECRET's predicted spectra, for random (representative) choices of the parameter set $\mathcal{P}= (\mathcal{R}_{\mathrm{SN}}[10^4\unit{Myr^{-1}}], \log_{10}(E_{\mathrm{cut}}[\unit{GeV}]), \gamma, \delta, \kappa_0[10^{28}\unit{GeV}])$. Individual samples from SECRET are plotted as coloured lines, the grey bands (black lines) are the $68\%$ and $90\%$ quantiles (medians) of the ensemble of samples. The hatched regions (dashed lines) show the quantile bands (medians) of the simulated dataset for comparison. For \textbf{(a)}-\textbf{(c)}, while not quite as accurate as in the much simpler single-point case, the predictions align with the true distributions. \textbf{(d)} shows one example where the predictions are completely off and unphysical. This happens at the very corner of the parameter space, see section \ref{sec:SECRET} for discussion.}
	\label{fig:SECRETSpectraExamples}
\end{figure}

In summary, the SECRET dataset contains $6750$ different parameter combinations with 19 energy bins + 5 parameter bins. For each point, $N_\mathrm{realisations} = 10^4$ individual realisations are simulated. This means that there is a factor of $10$ fewer datapoints per parameter point as for the single-scenario MADE, but the network also has additional information from surrounding datapoints, from which it can learn the general structure and correlations of spectra. In total, the final dataset contains $67.5$ million data points with 24 dimensions.



We train the model for 22 epochs, after which the loss starts to rise, which takes about $\mathcal{O}(1)\unit{d}$ on a GPU.

The trained network can generate stochastic spectra conditioned on the physical model parameters. We show some examples at random representative phase space points in Fig.~\ref{fig:SECRETSpectraExamples}. In most of the space, the learned distribution agrees fairly well with the simulations. However, there are some points at which the prediction fails completely, and non-physical spectra with huge fluctuations are generated.

To evaluate SECRET's performance more systematically, we need to define a summary statistic that quantifies the quality of the predictions in terms of the deviation to the simulations.
This is necessary to be able to assess the phase space coverage and to quantify in which regions the network fails. We calculate the difference of absolute intensity values at selected quantiles.

\begin{figure}[h]
	\begin{subfigure}{0.5\textwidth}
		\centering
		\includegraphics[width=\textwidth]{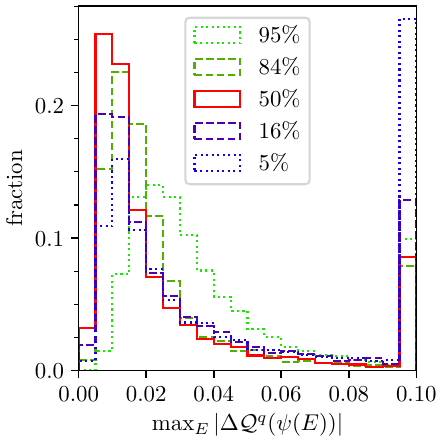}
		\caption{}
		\label{fig:SECRETQuantileDeviationMax}
	\end{subfigure}
	\hfill
	\begin{subfigure}{0.5\textwidth}
		\centering
		\includegraphics[width=\textwidth]{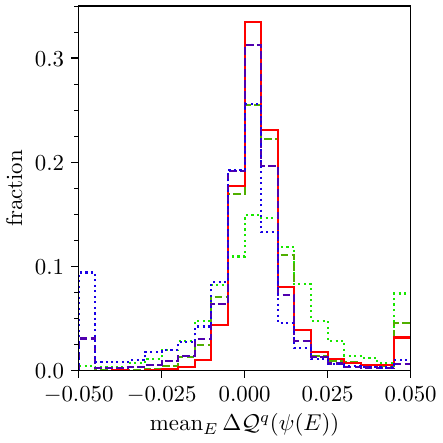}
		\caption{}
		\label{fig:SECRETQuantileDeviationMean}
	\end{subfigure}

	\caption{Distribution of $q$-quantile differences $\Delta\mathcal{Q}^q = \mathcal{Q}^q_\mathrm{SECRET} - \mathcal{Q}^q_\mathrm{sim}$ between dimensionless fluxes of SECRET predictions and simulations. The colours and linestyles denote different quantiles (corresponding to different values of $q$). The first and last bins are overflow bins. \textit{Left:} Maximum absolute deviation over energy bins. The distributions peak close to $0$, but do exhibit long tails of large errors. For the median and inner quantiles, $\sim90\%$ of parameter combinations have a maximum deviation much smaller than $0.1$, while the outer quantiles naturally exhibit larger errors. \textit{Right:} Average deviation over energy bins. The distributions peak slightly above 0 and are mostly contained within a small interval. For the median and inner quantiles, $\gtrsim90\%$ of values lie within $\pm0.025$. Again, the fraction of outliers is naturally higher for the outer quantiles.}
	\label{fig:SECRETQuantileDeviation}
\end{figure}

These absolute errors in the predictions are plotted in Fig.~\ref{fig:SECRETQuantileDeviation}. We see that the inner quantiles are predicted more accurately, while the 5\% quantiles in the tails show larger deviations. Overall, the errors are very close to zero for a large portion of the parameter space. Still, there are a number of outliers with large errors that still need to be investigated further.

\begin{figure}[h]
    \centering
	\begin{subfigure}{0.32\textwidth}
		\centering
		\includegraphics[width=\textwidth]{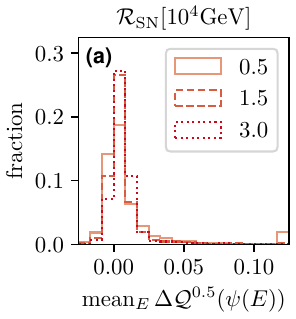}
		\phantomcaption{}
		\label{fig:SECRETQuantileDeviationPerSNrate}
	\end{subfigure}
	\hfill
	\begin{subfigure}{0.32\textwidth}
		\centering
		\includegraphics[width=\textwidth]{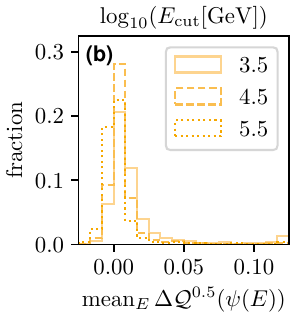}
		\phantomcaption{}
		\label{fig:SECRETQuantileDeviationPerEcut}
	\end{subfigure}
	\hfill
    \begin{subfigure}{0.32\textwidth}
		\centering
		\includegraphics[width=\textwidth]{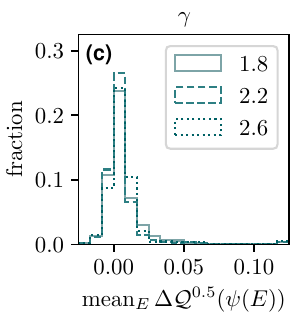}
		\phantomcaption{}
		\label{fig:SECRETQuantileDeviationPerGamma}
	\end{subfigure}

    \begin{subfigure}{0.32\textwidth}
		\centering
		\includegraphics[width=\textwidth]{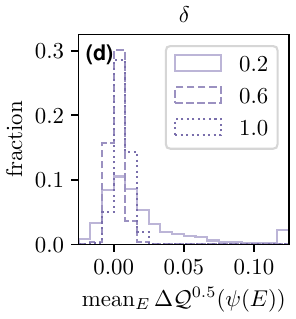}
		\phantomcaption{}
		\label{fig:SECRETQuantileDeviationPerDelta}
	\end{subfigure}
    \begin{subfigure}{0.32\textwidth}
		\centering
		\includegraphics[width=\textwidth]{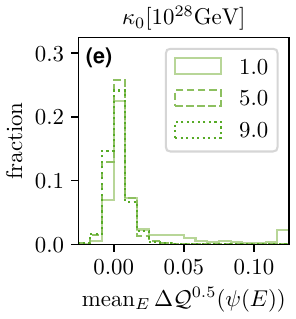}
		\phantomcaption{}
		\label{fig:SECRETQuantileDeviationPerD0}
	\end{subfigure}
	
	\caption{Histograms of mean deviations of the median $\mathcal{Q}^{0.5}$. Each plot shows a different parameter slice, with one parameter varied while the others are kept fixed at $\mathcal{R}_{\mathrm{SN}}=2\cdot10^4\unit{Myr^{-1}}$, $E_\mathrm{cut}=10^{4.5}\unit{GeV}$, $\gamma=2.2$, $\delta=0.6$, and $\kappa_0=5\cdot10^{28}\unit{cm^2s^{-1}}$. For the sake of clarity, only 3 values are shown: the middle and edges of the allowed ranges. The first and last bins are overflow bins.
    The distribution of errors does not significantly depend on $\gamma$. For $\mathcal{R}_\mathrm{SN}$ and $E_\mathrm{cut}$ there is a significant fraction of outliers only for their respective lowest values, while errors are well constrained for the rest of the parameter space. This effect is even stronger for the diffusion coefficient: mean deviations are always much smaller than $0.05$, except for the edge of the parameter space where the distribution significantly broadens and outliers appear. This implies that large errors only happen for specific parameter combinations, while in general deviations are mostly well contained in a small interval around 0.}
	\label{fig:SECRETQuantileDeviationSlices}
\end{figure}

To understand which conditioned parameter values are reliable and which lead to inaccurate results, Fig.~\ref{fig:SECRETQuantileDeviationSlices} shows the errors for specific values of the model parameters, where we varied one parameter at a time and marginalised over the rest. We see that indeed only specific values of certain parameters lead to a long tail of the distribution, namely only the smallest bins of the SN-rate and the diffusion coefficient parameters.

This allows us to identify these specific edges as responsible for the erroneous predictions we see e.g. in Fig.~\ref{fig:SECRETSpectraExamples}(d). This behavior is confirmed when inspecting the influence of pairwise parameter combinations, which we show in the appendix in Fig.~\ref{fig:SECRETQuantileDeviationCorner}.

Finally, this also implies the complementary statement: SECRET is reliably accurate on most of the parameter space, with a clearly defined area of validity. In fact, as Fig.~\ref{fig:SECRETQuantileDeviation} shows, the model is able to predict stochastic cosmic ray fluxes with few-percent accuracy. Alternatively we can formulate this as a guiding principle for the usage of SECRET: As long as one stays away from the lowest values of $\mathcal{R}_{\mathrm{SN}}$, $\delta$ and $\kappa_0$, samples are robust.

While this is true only on the points contained in the training set, i.e. only on the hypercubic grid we defined in Tab.~\ref{tab:ParamRanges}, we also tested the interpolation capabilities of the model. For that, we repeated the analysis on a new sample of simulations on a grid that is diagonally offset with respect to the training grid, that means on points within the original range with the maximum euclidean distance to the nearest training point. We find that even in that case, the errors are typically below $\Delta\mathcal{Q}^q(\psi(E)) \lesssim 0.05$, which corresponds to relative deviations of $\lesssim \mathcal{O}(10\%)$.

\subsection{Worked Example}

Along with this paper, we are releasing the entire source code for MADE and SECRET on GitLab\footnote{\url{https://git.rwth-aachen.de/pmertsch/secret}}.  The entire code-base is fully documented, and additionally contains a notebook with explanations and examples on how to use the models. For further details, please refer to the notebook or the documentation inside the code.

Additionally to the \verb|MADE| class, that implements the entire NN functionality, the repository also contains a class \verb|SECRET|, that implements the SECRET model architecture as it is described in this paper and handles everything from input standardisation to loading the pre-trained weights. With this, utilising SECRET to generate stochastic spectra is as easy as two lines of code:

\begin{verbatim}
    import SECRET-code

    secret = SECRET-code.models.SECRET()
    samples = secret.sample_by_par(n,SNrate,logEcut,gamma,delta,D0)

    SECRET-code.util.plot_fluxes(samples)
\end{verbatim}

Furthermore, the code also contains scripts to train MADE or SECRET from scratch, which can be executed e.g. as follows:

\begin{verbatim}
    python train_made.py
    -e <number of epochs to train for>
    -s <the epoch to start the training at, loads pre-trained model>
    -i <number of input nodes>
    -l <add a hidden layer with that many nodes, stacks>
    -o <number of output components>.
\end{verbatim}

For example, the call for the MADE training used as the single-point model in this paper looks like this: \verb|python train_made.py -e 500 -i 19 -l 200 -o 10|.

\section{Conclusions}
\label{sec:conclusions}

In this work we introduced SECRET, a machine learning model for fast generation of stochastic spectra, eliminating the need to run computationally expensive MC-simulations.
With the inherent autoregressive properties baked into the underlying architecture, MADE, the predicted distribution is conditioned on physical parameters such as the supernova rate, the source spectrum of CR electrons, and the diffusion coefficient.

We trained the model on a large 5D parameter space which covers a wide range of parameter values. Comparing the learned density to simulations we find that SECRET is able to predict the quantiles of the distribution to percent level, and we quantify the area of validity in which SECRET's predictions are reliable.
Future extensions should include the effects from regions of slow diffusion, inferred from the observations of so-called gamma-ray halos around middle-aged pulsar wind nebulae~\cite{HAWC:2017kbo,Fang:2022fof}.

The trained model can now predict the conditional pdf of CR electron spectra which serves two main purposes: It can evaluate exact likelihoods, which was previously not accessible due to the theory model being intractable. This enables further studies to constrain physical parameters in a simulation-based inference \cite{doi:10.1073/pnas.1912789117} setup.

Secondly, the model can generate new samples very efficiently, effectively serving as an emulator for the Monte Carlo simulations and alleviating the need for expensive runs. As such, it is now straightforward to get an ensemble of spectra including a direct estimate for the amplitude of stochastic variations that can be expected at a given energy bin under some assumed physics model.

MADE interpolates in-between the training grid and enables sampling of points that were not included in our dataset. However, it can not be expect to extrapolate outside of the trained range with any reliability.

The entire pre-trained network will be publicly available and free to use for emulation of stochastic effects in further studies. Additionally, we provide the code for MADE so this method of density estimation can be extended to other physical models by retraining on a new dataset.


\appendix

\section{Appendix}

\begin{figure}[h]
	\centering
	\includegraphics[width=0.95\textwidth]{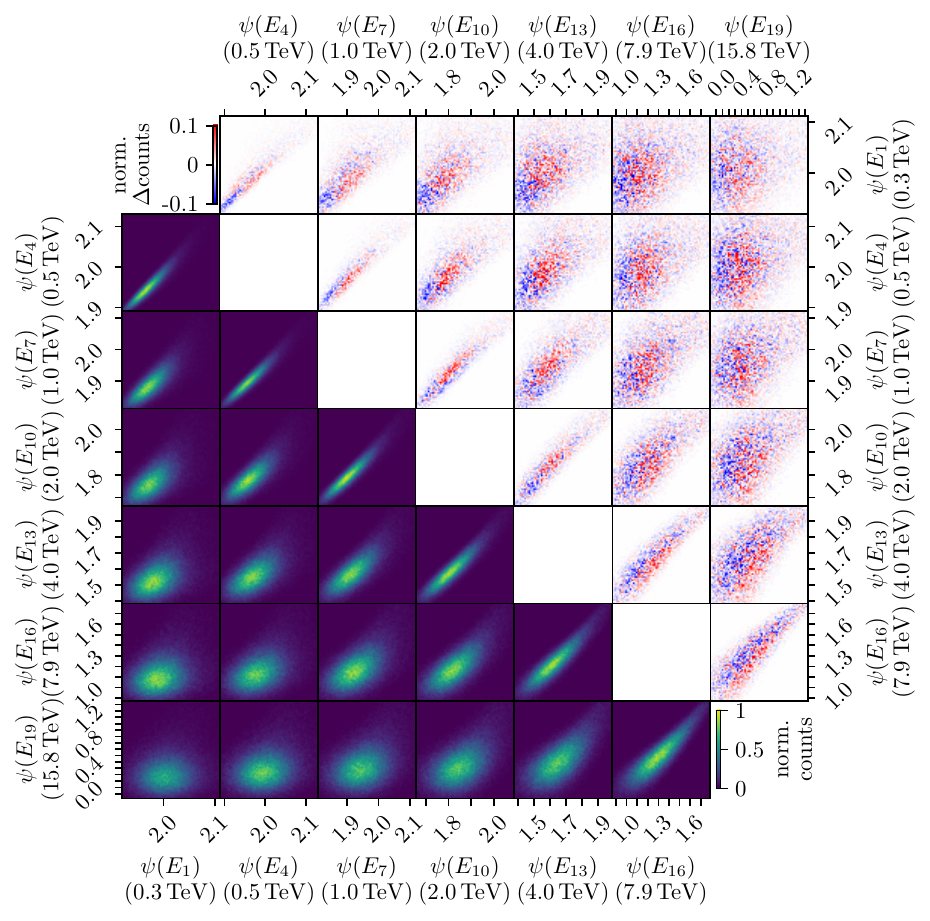}
	\vspace{-0.5cm}
	\caption{Pairwise marginalised histograms of dimensionless flux. For the sake of clarity, we only show every third bin. \textbf{Lower left:} MADE samples. Each histogram is normalised by dividing the counts per bin by the maximum counts in each respective subplot for better visibility. Bins that are more closely together are also more tightly correlated, while bins that are farther apart are less correlated, which aligns with our physical expectations. Additionally, the correlation structure is clearly non-linear, which demonstrates that the NN is able to capture the non-gaussian information. The distributions are visually identical to the target simulation ones, so we only show the normalised difference in the \textbf{upper right}: $\frac{counts_\mathrm{MADE}-counts_\mathrm{sim}}{\max(counts_\mathrm{MADE})}$. There is no strong bias overall and the slight discrepancies can be attributed to a lack of coverage consistent with figure \ref{fig:MADEPredictions1DHists}, which leads to MADE samples being slightly more concentrated than the simulations.}  
	\label{fig:MADEPredictionsCorner}
\end{figure}

\begin{figure}[h]
	\centering
	\includegraphics[width=\columnwidth]{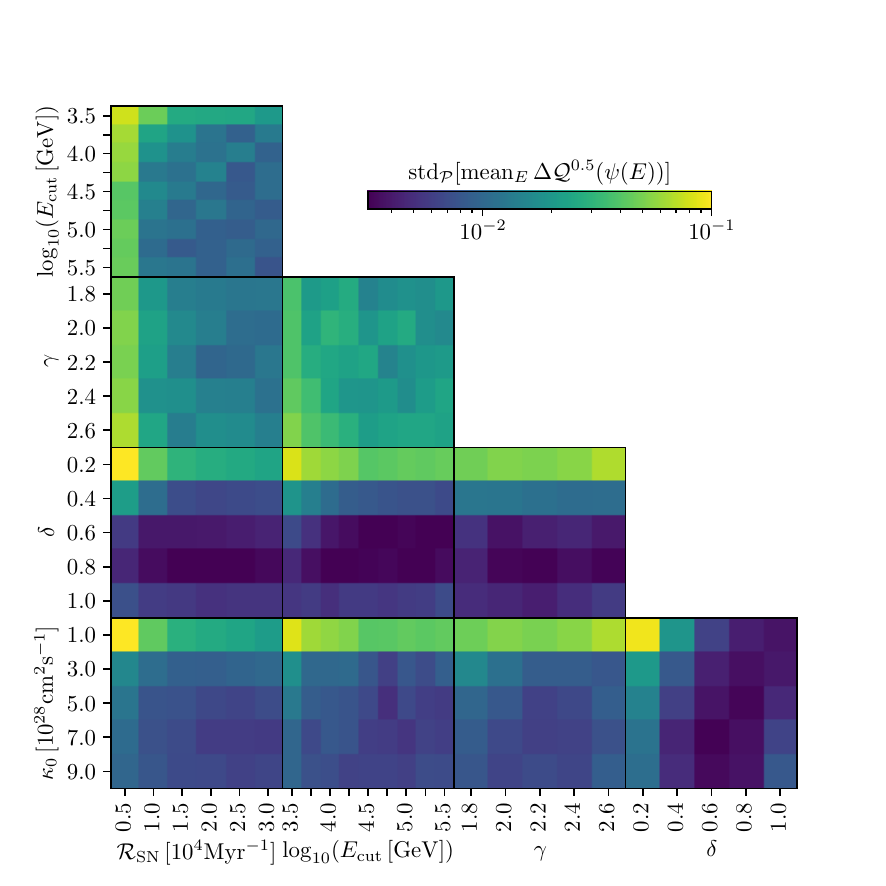}
	\caption{Pairwise marginalised standard deviations of mean differences in median. This corresponds the spread of the distributions shown in figure \ref{fig:SECRETQuantileDeviation} (except now conditioned on two parameters) and is thereby a measure of the error of SECRET's prediction. Large values on the color scale correspond to a wide distribution of the mean of $\Delta\mathcal{Q}^{0.5}$, i.e. a long tail/outliers in the deviation between SECRET and simulations.
    We see that large values are mostly constrained to specific parameter combinations. Specifically, the diffusion coefficient parameters have a large impact at the edge of the parmeter space, while exhibiting consistently low values elsewhere. Note that the relatively high minimum values in the upper three panels are explained by marginalising over the most extreme outliers caused by $\kappa_0$ and $\delta$. This supports the conclusion that the outliers observed in SECRET's predictions are well localised, implying that SECRET is reliable over the rest of it's trained range.}
	\label{fig:SECRETQuantileDeviationCorner}
\end{figure}

In the following, we provide some additional figures for characterising the performance of the MADE and SECRET models. 
In Fig.~\ref{fig:MADEPredictionsCorner}, we show the 2D marginalised distributions from MADE in comparison with the Monte Carlo simulations. 
Fig.~\ref{fig:SECRETQuantileDeviationCorner} reports the variance of mean median deviations for different combinations of model parameters. 

\clearpage
\bibliographystyle{JHEP}
\bibliography{bib}

\end{document}